\documentclass[aps, prl, reprint,
 amsmath,amssymb, amsfonts,
 floatfix,
]{revtex4-2}

\usepackage{graphicx}
\usepackage{dcolumn}
\usepackage{bm}
\usepackage{siunitx}
\usepackage{epstopdf}
\usepackage{tikz}
\usepackage{mathtools}

\DeclareSIUnit\nanomolar{nM}
\newcommand{\Dbar}{\overline{D}}
\newcommand{\kbar}{\overline{k}}
\newcommand{\cbar}{\overline{c}}
\newcommand{\vbar}{\overline{v}_1}
\newcommand{\xbar}{\overline{x}}
\newcommand{\tbar}{\overline{t}}
\newcommand{\rhoeq}{\rho_\infty}

\begin{document}

\title{Herding of proteins by the ends of shrinking polymers}

\author{Amer Al-Hiyasat}

\affiliation{
 Department of Molecular Biophysics and Biochemistry, Yale University, New Haven, Connecticut 06511, USA
}
\author{Yazgan Tuna}
\affiliation{
 Department of Molecular Biophysics and Biochemistry, Yale University, New Haven, Connecticut 06511, USA
}
\author{Yin-Wei Kuo}
\affiliation{
 Department of Molecular Biophysics and Biochemistry, Yale University, New Haven, Connecticut 06511, USA
}
\author{Jonathon Howard}
    \email{Correspondence: joe.howard@yale.edu}
\affiliation{
 Department of Molecular Biophysics and Biochemistry, Yale University, New Haven, Connecticut 06511, USA
}

\date{\today}

\begin{abstract}
The control of biopolymer length is mediated by proteins that localize to polymer ends and regulate polymerization dynamics. Several mechanisms have been proposed to achieve end localization. Here, we propose a novel mechanism by which a protein that binds to a shrinking polymer and slows its shrinkage will be spontaneously enriched at the shrinking end through a ``herding" effect. We formalize this process using both a lattice-gas model and a continuum description, and we present experimental evidence that the microtubule regulator spastin employs this mechanism. Our findings extend to more general problems involving diffusion within shrinking domains.
\end{abstract}
\maketitle
Regulating the length of biological polymers is essential for myriad cellular processes such as mitosis, muscle contraction and ciliary motility \cite{Marshall2015}. This regulation is often mediated by proteins that localize to polymer ends, where they promote or inhibit polymerization or depolymerization \cite{Howard2007}. The mechanisms of protein end-localization on growing polymers such as microtubules and actin filaments have been well documented, both experimentally and theoretically. Some proteins bind directly to ends from solution \cite{Bieling2007, Edwards2014}, some bind the polymer near the end and undergo one-dimensional diffusion to reach the end \cite{Reithmann2016,Helenius2006,Brouhard2008}, and others are carried to ends by motor proteins \cite{Varga2009, Klein2005, Chen2019}. Much less is known, however, about the localization of proteins to shrinking ends.    

	Microtubule severing proteins such as spastin and katanin cut microtubules along their lengths using the energy of ATP hydrolysis \cite{McNally2018}. In addition to severing, both spastin and katanin possess an ATP-independent activity that promotes “rescue”, the conversion of shrinking microtubule ends to growing ones. This activity leads to the amplification of microtubule mass \cite{Vemu2018, Kuo2019b}, which is critical to spastin’s biological function in cells  (reviewed in \cite{Kuo2021}). Furthermore, it has recently been reported that spastin is significantly enriched on shrinking microtubule ends, where it slows the shrinkage rate and facilitates the regrowth-promoting activity \cite{Kuo2019a}. The mechanism of this enrichment is unknown.
	
	End localization by spastin is not likely to involve direct binding or diffusion-and-capture: such mechanisms would require very high attachment rates and diffusion coefficients to target shrinking microtubule ends, which move 10-40-fold faster than growing ones \cite{Kuo2019a}. A potential mechanism is that spastin uses multivalent interactions (avidity) with the microtubule to maintain contact with the depolymerizing end, as is thought to occur for end-tracking multiprotein complexes. For example, NDC80 forms a large sleeve that ensheathes shrinking microtubules in the mitotic spindle \cite{Alushin2010}, the DASH/Dam1 complex forms a ring that encircles microtubules\cite{Westermann2006}, and Kar9 with associated proteins remains on the ends of shrinking cytoplasmic microtubules \cite{Meier2021}. Engineered \cite{Lombillo1995} and synthetic oligomers \cite{Drechsler2019} can also robustly track shrinking ends. While the avidity mechanism cannot be ruled out for spastin, whose hexameric form cuts microtubules, there is no evidence that spastin end-localization requires oligomerization. Indeed, spastin binds to and diffuses along microtubules as a monomer and localizes to their ends at low concentration and in the absence of ATP \cite{Kuo2019a}. 
	
	In this Letter, we present a novel theoretical mechanism of shrinking-end localization that requires neither multivalent interactions, diffusive capturing, nor directed motility. We show that a protein that binds to a shrinking polymer and slows its shrinkage will be spontaneously enriched at the shrinking end due to a purely kinetic ``herding" effect. This behavior is counter-intuitive because one might expect that slowdown is driven by end enrichment, not the other way around. We experimentally demonstrate that spastin’s end enrichment is quantitatively explained by this herding theory. In addition to providing insight into spastin’s regulatory function, our results extend to a more general class of problems in which diffusive particles hinder the boundaries of a shrinking domain.

\paragraph{Mathematical Model.} 

\begin{figure}
    \centering
    \begin{tikzpicture}
        \node  (image) at (0,0) {
        \includegraphics[scale=0.97]{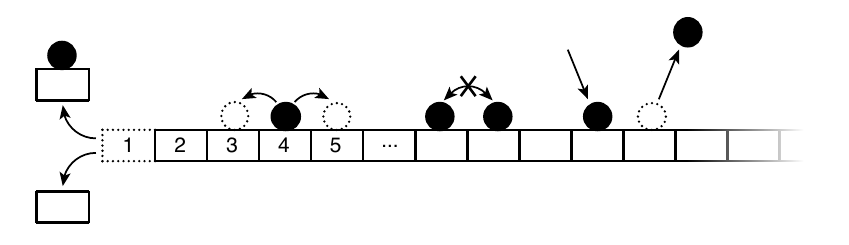}};
        \node at (1.15,.4) {$\omega_\mathrm{a}c$};
        \node at (2.7,.4) {$\omega_\mathrm{d}$};
        \node at (-0.95,.45) {$\omega_\mathrm{h}$};
        \node at (-1.75,.45) {$\omega_\mathrm{h}$};
        \node at (-3.76,.02) {$\omega_1$};
        \node at (-3.76,.-.43) {$\omega_0$};
        \end{tikzpicture}
    \caption{Microscopic model of protein dynamics on shrinking polymer. Proteins bind to the lattice with rate $\omega_\mathrm{a}$, detach with rate $\omega_\mathrm{d}$, and hop with rate $\omega_\mathrm{h}$. The proteins exclude each other from lattice sites. The site $i=1$ shrinks away at a rate $\omega_0$ if it is vacant and $\omega_1$ if it is occupied.}
    \label{fig:schematic}
\end{figure}

Lattice-gas models have been successfully employed to explain the collective dynamics of polymer-associated proteins \cite{Kolomeisky1998,Parmeggiani2003}, including depolymerizing kinesins \cite{Klein2005} and end-binding microtubule polymerases \cite{Reithmann2016}. In this vein, we describe the shrinking polymer as a semi-infinite one-dimensional lattice situated in a reservoir of the soluble protein (figure \ref{fig:schematic}). We index the lattice sites $i = 1,2,3...$ where $i = 1$ is the shrinking end. The lattice represents protein binding sites, which can bind only one protein at a time. Proteins from solution bind vacant lattice sites at a rate  $\omega_\mathrm{a} c$, where $c$ is the concentration of the protein in solution. Lattice-bound proteins can detach at a rate $\omega_\mathrm{d}$, or hop to adjacent vacant sites at a rate $\omega_\mathrm{h}$ (in either direction). Shrinkage of the polymer occurs via the loss of lattice sites from one end at a rate $\omega_0$ or $\omega_1$, depending on whether the site at the end is vacant or occupied, respectively. We are interested in the case $\omega_1 < \omega_0$, corresponding to a slowing of shrinkage by the protein. 

The state of the system is described by the set of occupancy numbers $\{n_i\}$ where $n_i =1$ if the $i$th site is occupied. Under these definitions, the time evolution of the mean occupancy at the end $\langle n_1 \rangle (t)$ obeys
\begin{eqnarray} \label{disc1}
 \frac{d \langle n_1 \rangle}{dt} = \,&& \omega_\mathrm{h} (\langle n_2 \rangle - \langle n_1 \rangle) + \omega_\mathrm{a} c \langle 1- n_1 \rangle - \omega_\mathrm{d} \langle n_1 \rangle \nonumber\\
 &&+ \omega_0 \langle  (1-n_1)(n_2-n_1) \rangle \nonumber \\
 &&+\omega_1 \langle n_1(n_2-n_1)\rangle.
\end{eqnarray}
Similarly, the mean occupancy of all other sites $i>1$ follows
\begin{eqnarray} \label{disci}
  \frac{d \langle n_i \rangle}{dt} = &&\omega_\mathrm{h} \left( \langle n_{i+1} \rangle - 2 \langle n_i \rangle + \langle n_{i-1} \rangle \right) + \omega_\mathrm{a} c \langle 1- n_i \rangle \nonumber\\
  &&- \omega_\mathrm{d} \langle n_i \rangle \nonumber
  + \omega_0 \langle (1-n_1)(n_{i+1}-n_i) \rangle \\&&+ \omega_1 \langle n_1 (n_{i+1}-n_i)\rangle.
\end{eqnarray}
Equations \ref{disc1} and \ref{disci} are derived in section I of the Supplementary Information (SI). Due to the moving frame, there is an apparent advection of particles towards the shrinking end at $i = 1$.

In the case $\omega_0 =\omega_1$ (no slowdown by end-occupancy), equations \ref{disc1} and \ref{disci} have the stationary solution
\[\langle n_i \rangle = \frac{\omega_\mathrm{a} c}{\omega_\mathrm{a} c + \omega_\mathrm{d}} = \rhoeq\]
for all $i$. This is the \textit{Langmuir isotherm}, the stationary density that would be achieved if the polymer was in equilibrium with the protein bath and constant in length. For $\omega_0 \neq \omega_1$, direct solution of equations \ref{disc1} and \ref{disci} is complicated by the presence of correlations between $n_1$ and the other sites, but we can invoke a mean field (MF) approximation $\langle n_1 n_i \rangle \approx \langle n_1 \rangle \langle n_i \rangle$. The continuum limit of the lattice model is then
\begin{eqnarray} 
    \partial_{\tbar} \rho(\xbar,\tbar) =&& \Dbar \, \partial_{\xbar}^2 \rho(\xbar,\tbar) + [1-(1-\vbar) \rho(0,\tbar)]\, \partial_{\xbar} \rho(\xbar,\tbar) \nonumber\\ &&+ \kbar \cbar (1 - \rho(\xbar, \tbar)) - \kbar \rho(\xbar, \tbar),  \label{pde}
\end{eqnarray}
where we have defined the continuous spatial variable $\xbar = i-1$ as well as the dimensionless time $\tbar = t \omega_0$, which normalizes the natural polymer shrinkage rate to unity. The mean linear density of protein on the polymer is $\rho(\xbar,\tbar) = \langle n_{(x+1)} \rangle (\tbar/\omega_0)$. We further define $\Dbar = \omega_\mathrm{h}/\omega_0$,  the ratio of the hopping rate to the natural shrinkage rate,  $\kbar = \omega_\mathrm{d} / \omega_0$, the dimensionless detachment rate of the protein, and  $\cbar = \omega_\mathrm{a} c/\omega_\mathrm{d}$, which determines the position of Langmuir equilibrium $\rhoeq = \frac{\cbar}{1+\cbar}$. Lastly, we define a slowdown parameter $\vbar = \omega_1/\omega_0$ controlling the extent to which the protein slows shrinkage. When  $\vbar=0$, depolymerization is only possible when the terminal site is vacant, and when  $\vbar=1$, the depolymerization rate is unaffected by end occupancy. From equation \ref{pde}, we identify the (dimensionless) measured shrinkage velocity of the microtubule
\[\overline{v}(\tbar) = 1-(1-\vbar) \rho(0,\tbar).\]
\begin{figure*}
    \centering
    \includegraphics{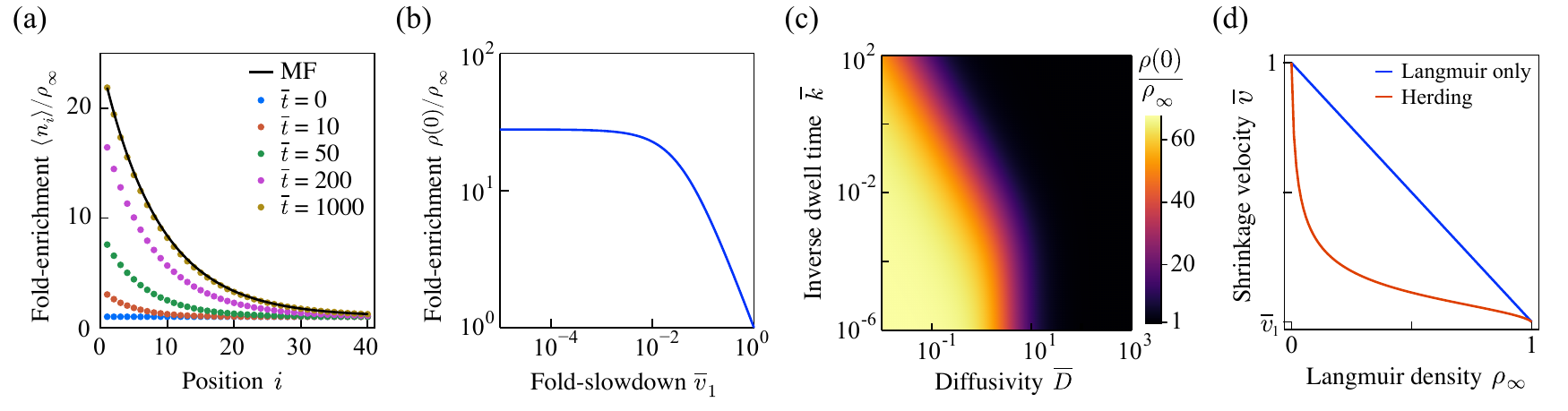}
    \caption{Herding theory of protein enrichment at shrinking ends. (a) Occupancy probability at each position at different times. Circles: Estimates from simulations of the lattice-gas model. Line: mean-field solution. (b) Fold enrichment at the end as a function of $\bar{v}_1 = \omega_1/\omega_0$. (c) Fold enrichment at the end as a function of $\Dbar$ and $\kbar$. (d) Mean microtubule shrinkage velocity as a function of the equilibrium protein density. }
    \label{fig:theory}
\end{figure*}
Far from the shrinking end, the density of the protein should be the equilibrium Langmuir density $\rhoeq$. With this boundary condition, stationary solutions to equation \ref{pde} take the form of an exponential decay with the distance from the end,
\begin{equation} \label{statSol}
    \rho(\xbar) = (\rho(0)-\rhoeq) e^{-\lambda \xbar} + \rhoeq,
\end{equation}
where 
\[\lambda = \frac{1\!-\!(1\!-\!\vbar)\rho (0) \!+\! \sqrt{4 \Dbar \,\kbar(1\!+\!\cbar)+\left[1\!-\!(1\!-\!\vbar)\rho (0)\right]^2}}{2 \Dbar}.\]
To set the boundary condition at the end, we note that the microscopic model has no binding sites to the left of $\xbar = 0$, and so the end must be reflecting. This gives
\begin{equation} \label{endbound}
-\Dbar\, \partial_{\xbar}\rho(0) - \overline{v} \rho(0, \tbar) =  \kbar \cbar (1 - \rho(0)) - (\kbar + \vbar) \rho(0),
\end{equation}
where we have imposed that the total flux (diffusive and advective) through the end must equal the net binding or detachment at the end. This condition can be used together with equation \ref{statSol} to express $\rho(0)$ as the root of a cubic polynomial (SI, section II). Notably, the solutions predict $\rho(0) > \rhoeq$, implying that there is an enrichment of the protein at the shrinking filament end. Simulations of the microscopic model using the Gillespie algorithm predict that the mean lattice occupancies approach a steady state that is in good agreement with the stationary solution of the mean field model (figure \ref{fig:theory}a). This agreement is robust against changes in the parameters (figure S1).

The gradual accumulation at the end is due to a ``herding" effect, where the shrinking end slows when it encounters a protein and then speeds up again when the protein hops away from the end along the lattice. This results in a build up of the protein near the reflecting end. This unexpected behavior, that slowdown causes end enrichment and not vice versa, depends sharply on the extent of slowdown: the enrichment is maximum when $\vbar =\omega_1=0$ (figure \ref{fig:theory}b); that is, when a protein in the $i=1$ site completely pauses depolymerization. Since every lattice site is indistinguishable to the protein, the herding is a purely kinetic effect and is not due to a higher affinity to the polymer end. If $\Dbar$ or $\kbar$ are made too large, the proteins achieve diffusive or Langmuir equilibrium between lattice depolymerization events, preventing end enrichment. That is, no build up at the end would be possible if the protein were to hop much faster than the end moves or if the lifetime of the protein on the polymer were small compared to the time between depolymerization events (figure \ref{fig:theory}c). This is in contrast to diffusion-and-capture mechanisms, in which a large diffusion coefficient increases the end localization rather than decreasing it \cite{Reithmann2016}, or to direct-binding mechanisms, in which fast lattice turnover promotes end enrichment.

Herding has two biological implications. First, far less protein is needed on the lattice in order to impart a certain level of shrinkage slowdown (figure \ref{fig:theory}d). This is due to a feedback effect where slowdown causes end enrichment and end enrichment in turn promotes slowdown. Second, a high end occupancy is achievable at low solution concentrations. This can facilitate the regulation of polymer ends through mechanisms additional to the slowdown itself, such as the promotion of rescue by spastin.

\begin{figure}
    \centering
    \includegraphics[scale=.99]{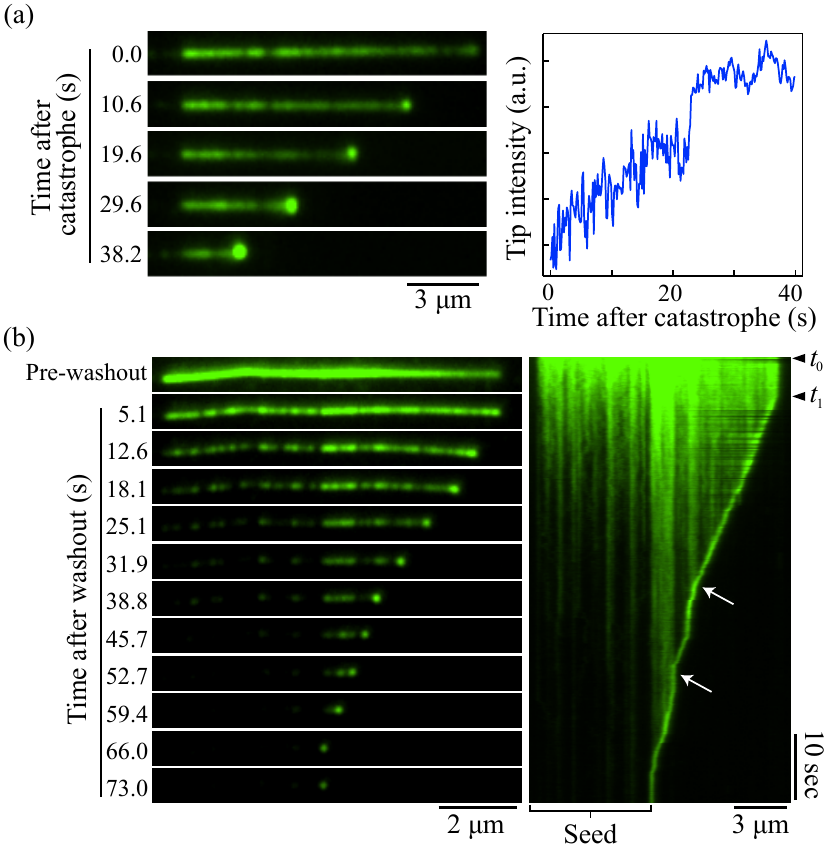}
    \caption{Experimental visualization of spastin on shrinking microtubules. (a) end accumulation of GFP-spastin was observed via TIRF microscopy. Left: representative time series. Right: end intensity trace generated by integrating a 1\textmu $\text{m}^2$ region centered at the end. (b) Spastin washout assay. Dynamic microtubules were incubated with GFP-spastin and tubulin, and then both spastin and tubulin were washed out. end enrichment was still observed in the time series (left) and kymograph (right). $t_0$ is the washout time and $t_1$ is the catastrophe time. Arrows indicate slowdown of the end.}
    \label{fig:experiment}
\end{figure}

\paragraph{Application to spastin.} To validate this model, we investigated the behavior of spastin labeled with green fluorescent protein (GFP-spastin) on dynamic microtubules using the in vitro reconstitution assay described in \cite{krismethods}. Experimental details are provided in section IV of the SI. Briefly: We grew GMPCPP-stabilized microtubule ``seeds", which do not shrink, and immobilized them on a functionalized surface within a microfluidic flow chamber. We perfused a tubulin and GTP solution, which causes dynamic microtubule extensions to grow from the ends of the seeds. We then perfused a tubulin, GTP, and spastin solution, allowing spastin to decorate the dynamic microtubules and reach its equilibrium binding density. To prevent severing, no ATP was included in the solution. Finally, we perfused a solution containing spastin but no tubulin. The removal of tubulin induces ``catastrophe" events, forcing the microtubule ends to begin shrinking. GFP-spastin was visualized using total-internal-reflection fluorescence (TIRF)-microscopy \cite{Tuna2022}. In all microtubules observed ($n>200$), robust end enrichment was evident, as seen in the sample time series and intensity traces in figure \ref{fig:experiment}a. The enriched region coincides with the position of the microtubule end, and occurs on both ends of a shrinking microtubule (figure S2). This experiment confirms that spastin concentrates on depolymerizing microtubule ends.

We then asked whether the end enrichment is due to spastin preferentially binding to the shrinking end from solution or if it is due to the sweeping of lattice-bound spastin. To this end, we employed a ``wash-out" assay, where spastin was loaded onto microtubules in the presence of tubulin, and then both tubulin and spastin were washed out. This induces catastrophe within a spastin-free solution. We found that spastin was still enriched at shrinking ends, as seen in the representative time series in figure \ref{fig:experiment}b (left), and in the corresponding kymograph (right). This result conclusively demonstrates that a direct-binding EB1-like mechanism \cite{Bieling2007} is not necessary for end enrichment. Indicated on the kymograph (white arrows) are points where the end slows down when encountering regions of high spastin density and then speeds up again when spastin detaches from the end. This indicates a correlation between end occupancy and shrinkage speed, supporting the herding effect. These results show that the enrichment of spastin on shrinking ends does not require spastin in solution and provide direct evidence that the shrinking end sweeps spastin with it as it moves. 

To quantitatively validate our theory, we directly measured spastin's attachment rate $k_\mathrm{a}$, detachment rate $k_\mathrm{d}$, and diffusion coefficient $D$. The conversion between the macroscopic experimental parameters ($D, k_\mathrm{a}$, and $k_\mathrm{d}$) and the model parameters ($\Dbar, \cbar,$ and $\kbar$) is given in Table S2 (SI). The attachment rate was estimated by counting the frequency of single-molecule landing events at \SI{10}{\nanomolar} GFP-spastin. This yielded the estimate $k_\mathrm{a} = \SI{0.031\pm 0.002}{\per \micro\meter \per \nanomolar \per \minute}$ (mean $\pm$ SEM; 153 events). To measure the diffusion coefficient and detachment rate, we performed a washout assay using a low initial spastin concentration to visualize single-molecule trajectories (figure S3). The spastin dwell time displayed the expected exponential distribution (figure S4a), with an estimated mean $1/k_\mathrm{d} =\SI{17.7 \pm 1.4}{\second}$ (158 events).  Diffusion coefficients were estimated on a per-trajectory basis and fell into a broad range (figure S4b), with a median of $D = \SI{0.008}{\micro\meter\squared\per\second}$ (interquartile range: 0.008).  These parameter values are similar to those measured for human spastin \cite{Eckert2012}.

\begin{figure}
    \centering
    \includegraphics[scale=.95]{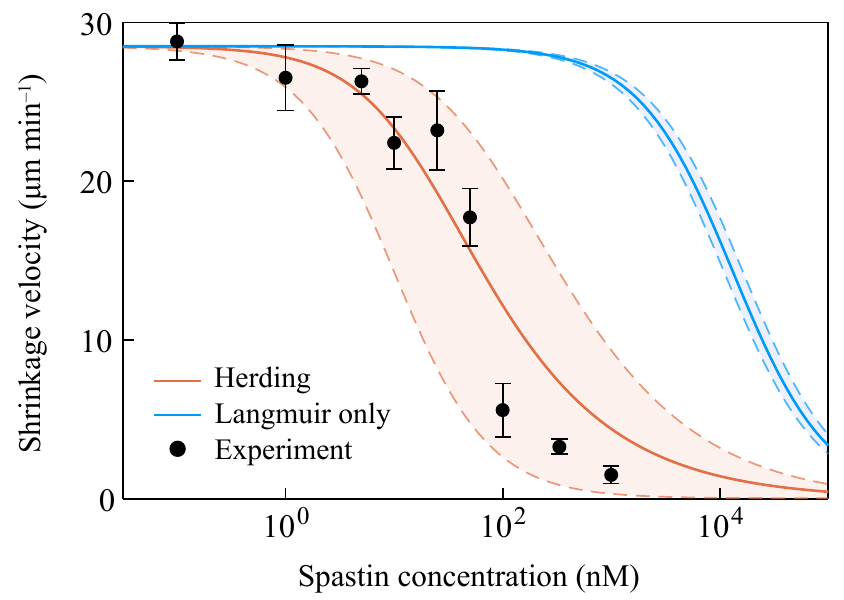}
    \caption{Microtubule shrinkage velocity as a function of spastin concentration. Circles: average estimate from at least 3 independent experiments (error bars: SD). 30-100 MTs were observed in each experiment by IRM. Orange line: prediction of the herding model. Blue line: prediction if $\rho(0) = \rho_\infty$. $v_0$ was estimated from measurements with no spastin. $v_1$ is set to zero. Shaded regions represent 95\% confidence intervals on $\Dbar$, $\kbar$ and $\cbar$, which were estimated from direct measurements. }
    \label{fig:shrinkvel}
\end{figure}

We next measured the microtubule shrinkage velocity $v$ as a function of the concentration of spastin in solution $c$ (figure \ref{fig:shrinkvel}, circles). The $v(c)$ curve allows an estimation of the slowdown parameter $\bar{v}_1$ as the ratio $v(\infty)/v(0)$, where $v(\infty)$ is the saturating shrinkage velocity at high concentrations. Direct measurement of $v(\infty)$ is complicated by spastin's tendency to aggregate at high concentrations. However, using the measurement at 1000 nM, we can place an upper bound at $\bar{v}_1 = 0.04$. Curve fitting suggests that the true value is even closer to zero. Thus, we were able to measure the key parameters in the model.

To make model predictions, we calculated the dimensionless parameters: $\kbar = 0.0010 \pm 0.0002$, $\Dbar \leq 2.0 \pm 0.5$, and $\bar{v}_1 \leq 0.04$. This places spastin in a region where herding is expected to be significant (figure \ref{fig:theory}b and c). At 50 nM, which is close to the physiological concentration \cite{Solowska2008, Itzhak2016}, the enrichment at the end is at least 21-fold and likely closer to 200-fold. The width of the enriched region is on the order $1/\lambda = \SI{75}{\nano\meter}$, which is below the diffraction limit and in agreement with experiments.

To compare the predictions directly to the data, figure \ref{fig:shrinkvel} shows the measured $v(c)$ values along with theoretical curves from our herding theory and for the case $\rho(0) = \rho_\infty$ (Langmuir kinetics without end enrichment). The shaded regions represent 95\% confidence intervals for the parameter estimates. It is evident that Langmuir kinetics alone are insufficient to explain the measured slowdown in the shrinkage velocity, even for $\bar{v}_1=0$. This suggests that spastin's end-enrichment is likely to be biologically significant, in that it allows slowdown at much lower cellular concentrations. According to experiments, a 50\% slowdown in the shrinkage velocity is achieved at around 55nM, which is close to the physiological concentration \cite{Solowska2008, Itzhak2016}. Without end-enrichment, this level of slowdown would be achievable only at micromolar concentrations (figure \ref{fig:shrinkvel}, blue curve). The experimental data are all within the confidence bounds of the herding theory (orange region), and it is clear that the theory is a good quantitative predictor of the measured shrinkage velocity.

\begin{figure}
    \centering
    \includegraphics[scale=1.0]{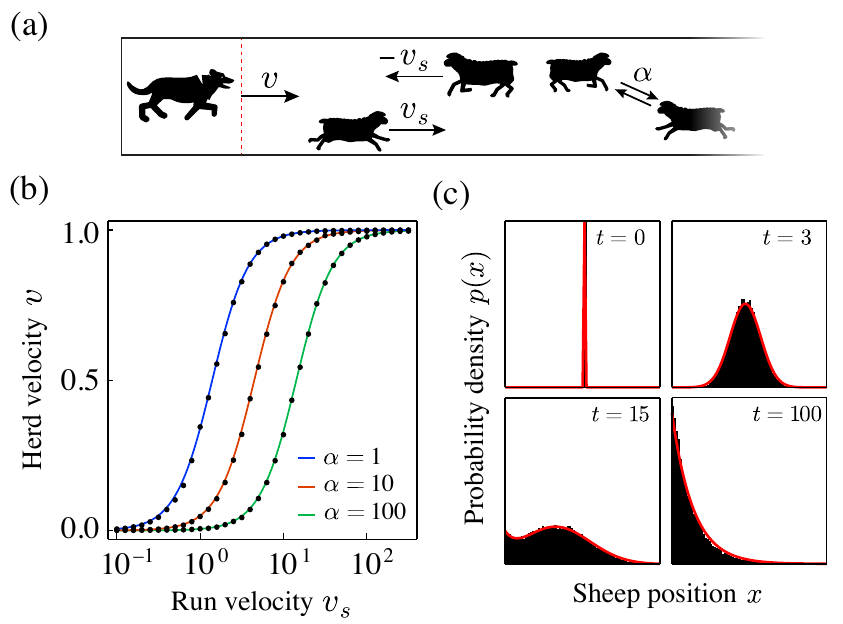}
    \caption{Toy model of sheepherding. (a) Sheep are noninteracting run-and-tumble particles that are reflected by the sheepdog. The sheepdog moves with unit velocity but pauses when it nears a sheep. (b) The average herd velocity $v$ as a function of the run velocity $v_\mathrm{s}$ and tumble rate $\alpha$. $N$ is equal to 5. Circles: Monte Carlo estimates from stochastic simulations. Lines: prediction of equation \ref{sheepsol}. (c) Sheep position distribution at different times. Black bars: simulated histograms ($N=1$). Red line: numerical Green's function of equation \ref{sheepsol}.}
    \label{fig:sheep}
\end{figure}

\paragraph{Other applications.} 
Although we have framed the herding theory in the context of regulatory proteins on shrinking polymers, the central principle is more general: particles that diffuse in a shrinking domain will enrich at the shrinking boundary if they hinder its shrinkage. The essence of this effect is contained within the following simplified version of the mean-field model,
\begin{equation} \label{simplified}
\partial_t \rho = [1-\rho(0,t)]\partial_x \rho + D \partial_x^2\rho 
\end{equation}
where the Langmuir kinetics are removed and $\vbar$ is set to zero. Equation \ref{simplified} serves as a generic theory of herding phenomena, independently of the specific microscopic model from which it was derived. To illustrate this, and to motivate our terminology, we consider the following toy model of sheepherding (figure \ref{fig:sheep}a): a one-dimensional domain is bounded on one end by a sheepdog, which moves with unit velocity in the positive direction when it is unimpeded. There are $N$ sheep in the domain, each behaving as an independent run-and-tumble particle with run velocity $v_\mathrm{s}$ and tumble rate $\alpha$. The sheep are reflected by the sheepdog, and the sheepdog pauses when there is a sheep within one unit of length ahead of it.

The problem of determining the resultant average velocity $v$ of the sheepdog (and the herd) appears complicated at first glance. However, numeric simulations suggest that the behavior is well predicted by equation \ref{simplified} with $D = v_\mathrm{s}^2/2\alpha$, which is the known effective diffusion coefficient for a run-and-tumble particle. With a reflecting boundary at $x=0$ and the condition $\rho(\infty,t) = 0$, the stationary solution to equation \ref{simplified} predicts a steady-state herd velocity of
\begin{equation} \label{sheepsol}
v = \frac{v_\mathrm{s}^2}{v_\mathrm{s}^2+2N\alpha}. \end{equation}
This result is in good agreement with simulations (figure \ref{fig:sheep}b). The dynamics of this process are also well-described by equation \ref{simplified}: Although time-dependent analytic solution is difficult due to the nonlocal nonlinearity in the advection term, numeric integration is straightforward with a nonlocal finite-difference scheme (SI, section III). This reveals that equation \ref{simplified} is a nearly exact Fokker-Planck equation for the sheep position when $N=1$ (figure \ref{fig:sheep}c).

\paragraph{Discussion.} 
 We have shown that a protein that slows a polymer's shrinkage will enrich near the shrinking end, leading to a feedback between slowdown and end enrichment. In our model, the binding sites on the polymer are energetically equivalent, and the equilibrium density is uniform across the lattice. What drives the density distribution out of equilibrium is the polymer’s shrinkage (figure \ref{fig:schematic}), which breaks detailed balance and is a dissipative process. For microtubules, the energy source is GTP hydrolysis, for actin filaments it is ATP hydrolysis \cite{Wang1985} and for exonuclease-mediate DNA depolymerization it is nucleotide triphosphate hydrolysis. It is known that such processes can generate mechanical work \cite{Koshland1988, Coue1991}; here, we show that they can also build protein density gradients with regulatory functions.

A central aspect of our model is that the build up near the end is due to a sweeping of lattice-bound protein. Although many MAPs diffuse on MTs and are reflected by MT ends, most fall off with depolymerizing subunits and are not swept by the end. Examples include tau \cite{Hinrichs2012,Castle2020} and kinesins 8 and 13 \cite{Leong2020}. Sweeping requires a mechanism that stops proteins at the end from falling off with the subunits that they occupy. One possibility is to processively track the end as a multivalent complex \cite{Volkov2018, Grishchuk2017, Meier2021}. Processive tracking has been explored theoretically by Klein et al. in the context of depolymerizing kinesin motors \cite{Klein2005}: using a formalism similar to ours, it was shown that accumulation at shrinking ends occurs if there is a sufficient probability that a protein at the end moves onto the next site when the terminal site dissociates. Our herding theory provides an alternative and distinct mechanism for sweeping – by slowing depolymerization when the end is occupied, proteins are not lost with depolymerizing subunits. This means that the protein can be swept without any processivity. 

We applied our herding theory to spastin’s end enrichment and slowing of depolymerization. Our experimental results show that the shrinking microtubule end sweeps lattice-bound spastin, eliminating direct end-binding as an alternative mechanism (figure \ref{fig:experiment}). We directly measured spastin's diffusion coefficient, dwell time, and slowdown activity, and determined that they imply significant herding (figure \ref{fig:theory}b and c). The herding theory quantitatively predicts the microtubule shrinkage rate as a function of spastin concentration; it explains how significant slowdown can be achieved at concentrations of 10 nM despite the average lattice occupancy being very low (figure \ref{fig:shrinkvel}). Although it is difficult to rule out processive tracking by oligomers as an alternative mechanism, we note that processive motion by single puncta was never observed in our single-molecule measurements. We further note that oligomerization and herding are not mutually exclusive; in fact, if oligomers slow depolymerization, then this will lead to herding.

Our findings suggest that spastin's slowdown activity is self-catalyzing because it concentrates the protein at the shrinking end. This effect is what allows regulation of the microtubule shrinkage velocity at spastin concentrations within the physiological range. Furthermore, herding has implications for spastin’s regulation of the rescue frequency: although the precise mechanism of spastin’s ATP-independent promotion of rescue is unclear, it is reasonable that the activity depends on an interaction with the end. For example, if spastin accelerates the binding of GTP-tubulin to the shrinking end, then its accumulation at the end can restore the GTP cap and thus the growing of that end.  

Lastly, we showed how a simplified version of our mean field model, Equation \ref{simplified}, can describe more macroscopic herding phenomena. The dynamics described by equation \ref{simplified} may arise in diverse contexts. An example from biology is the motion of molecular motors under the hindrance of diffusive roadblocks. This includes microtubule motors slowed by other MAPs \cite{Korten2008}, as well DNA replication forks and RNA-polymerases moving against DNA-binding proteins \cite{Epshtein2003, Weaver2019}. Outside of biology, such dynamics could arise in transport processes where a density buildup at the boundary slows the flow in the bulk. For example, the flow of a particle suspension through a filter.
\bibliography{bibliography}
\end{document}


\title{Supplementary Information: Herding of proteins by the ends of shrinking polymers}

\author{Amer Al-Hiyasat}
\author{Yazgan Tuna}
\author{Yin-Wei Kuo}
\author{Jonathon Howard}
\affiliation{
 Department of Molecular Biophysics and Biochemistry, Yale University, New Haven, Connecticut 06511, USA
}

\date{\today}
\vspace{100pt}
\maketitle

\tableofcontents 
\section{Analysis of the Stochastic Lattice-Gas Model} \label{analyticdisc}
We begin with the stochastic lattice-gas model defined in Figure 1 of the main text. The model describes a Markov jump process $\{ \mathbf{N}(t) = (n_1(t), n_2(t) , \dots) : t\geq0\}$ on a countable state space $\mathcal{S} = \{0,1\}^\infty$. The master equation of this process describes the time evolution of $\prob{\mathbf{N}(t) = \mathbf{n}}$ for every $\mathbf{n} \in \mathcal{S}$. Although this master equation can be constructed, its direct solution is not straightforward. It is easier to begin with equations of motion for $\prob{n_i(t)=1}$ at each $i$ separately. We are interested in
\[
     \frac{d}{dt} \prob{n_i(t) = 1} = \frac{d}{dt} \expect{n_i},
\]
where $\expect{X}$ denotes the expectation of $X$ at time $t$. First, we consider the case $i\geq 2$. We can condition on the four variables $(n_1, n_{i-1}, n_i, n_{i+1})$ and invoke the law of total expectations to write
\begin{equation}\label{expectimedep} \frac{d}{dt} \expect{n_i} = \frac{d}{dt} \expect{\expect{n_i| n_1, n_{i-1}, n_i, n_{i+1}}}  = \expect{\frac{d}{dt} \expect{n_i| n_1, n_{i-1}, n_i, n_{i+1}}}, \end{equation}
where the outer expectation is taken over the joint distribution of ($n_1, n_{i-1}, n_i, n_{i+1}$) at time $t$. Define the function
\[f(n_1, n_{i-1}, n_i, n_{i+1}) =  \frac{d}{dt} \expect{n_i| n_1, n_{i-1}, n_i, n_{i+1}} = \frac{d}{dt} \prob{n_i(t)=1| n_1, n_{i-1}, n_i, n_{i+1}}\]

For $n_i = 0$, this function gives the instantaneous rate of the jump $n_i=0 \to n_i=1$, and for $n_i = 1$, it gives minus the rate of the jump $n_i=1 \to n_i=0$. According to the definition of the model, these jump rates are fully specified by $n_1, n_{i-1}$, and $n_{i}$. This means that for fixed ($n_1, n_{i-1}, n_i, n_{i+1}$), $f$ is a deterministic function, and its values can be tabulated as in Table \ref{tab:enumeration} (left).
\begin{table}
    \centering
    \begin{tabular}{@{\quad}c@{\qquad}c@{\quad}}
    \hline\hline
    \quad & \quad \\[-5pt]
         $(n_1, n_{i-1}, n_i, n_{i+1})$ & 
         $f(n_1, n_{i-1}, n_i, n_{i+1}) $ \\[5pt] 
        
        \colrule
        $(0,0,0,0)$ & $\omega_\mathrm{a} c$ \\
        $(1,0,0,0)$ & $\omega_\mathrm{a} c$ \\
        $(0,1,0,0)$ & $\omega_\mathrm{h}$ \\
        $(0,0,1,0)$ & $-\omega_\mathrm{d}-2\omega_\mathrm{h}-\omega_0$ \\
        $(0,0,0,1)$ & $\omega_\mathrm{a} c+\omega_\mathrm{h}+\omega_0$ \\
        $(1,1,0,0)$ & $\omega_\mathrm{a} c+\omega_\mathrm{h}$ \\
        $(1,0,1,0)$ & $-\omega_\mathrm{d}-2\omega_\mathrm{h}-\omega_1$ \\
        $(1,0,0,1)$ & $\omega_\mathrm{a} c+\omega_\mathrm{h}+\omega_1$ \\
        $(0,1,1,0)$ & $-\omega_\mathrm{d}-\omega_\mathrm{h}-\omega_0$ \\
        $(0,1,0,1)$ & $\omega_\mathrm{a} c+2\omega_\mathrm{h}+\omega_0$ \\
        $(0,0,1,1)$ & $-\omega_\mathrm{d}-\omega_\mathrm{h}$ \\
        $(1,1,1,0)$ & $-\omega_\mathrm{d}-\omega_\mathrm{h}-\omega_1$ \\
        $(1,1,0,1)$ & $\omega_\mathrm{a} c+2\omega_\mathrm{h}+\omega_1$ \\
        $(1,0,1,1)$ & $-\omega_\mathrm{d}-\omega_\mathrm{h}$ \\
        $(0,1,1,1)$ & $-\omega_\mathrm{d}$ \\
        $(1,1,1,1)$ & $-\omega_\mathrm{d}$ \\
        \hline\hline
    \end{tabular} \quad
    \begin{tabular}{@{\quad}c@{\qquad}c@{\quad}}
    \hline\hline
    \quad & \quad \\[-5pt]
         $(n_1, n_2)$ & 
         $g(n_1, n_{2}) $ \\[5pt] 
        
        \colrule
        $(0,0)$ & $\omega_\mathrm{a} c$ \\
        $(1,0)$ & $-\omega_\mathrm{d} -\omega_\mathrm{h} - \omega_1$\\
        $(0,1)$ & $\omega_\mathrm{a} c + \omega_\mathrm{h} + \omega_0$ \\
        $(1,1)$ & $-\omega_\mathrm{d}$ \\

        \hline\hline
    \end{tabular}
    
    \caption{Tabulated values of $f$ (left) and $g$ (right). For $i>1$, $|f(n_1, n_{i-1}, n_i, n_{i+1})|$ gives the instantaneous switch rate of $n_i$ conditional on $(n_1, n_{i-1}, n_i, n_{i+1})$. Similarly, $|g(n_1, n_2)|$ gives the switch rate of $n_1$ conditional on $(n_1,n_2)$.}
    \label{tab:enumeration}
\end{table}
It can be verified from the table that the following equality holds
\begin{eqnarray*}
f(n_1, n_{i-1}, n_i, n_{i+1}) =&&  \omega_\mathrm{a} c (1- n_i)  - \omega_\mathrm{d}  n_i + \omega_\mathrm{h} \left(n_{i+1} - 2 n_i  + n_{i-1}\right) \nonumber\\
 +&& \omega_0 (1-n_1)(n_{i+1}-n_i) + \omega_1 n_1 (n_{i+1}-n_i). 
 \end{eqnarray*}
for all $(n_1, n_{i-1}, n_i, n_{i+1}) \in \{0,1\}^4$. Substituting this result into equation \ref{expectimedep}, we recover the expression given in the main text for $i \geq 2$
\begin{eqnarray} \label{disc}
\frac{d}{dt}\expect{n_i} =\,\, \mathbb{E}_t  [\,\omega_\mathrm{a} c &&(1- n_i)  - \omega_\mathrm{d}  n_i + \omega_\mathrm{h} \left(n_{i+1} - 2 n_i  + n_{i-1}\right) \nonumber\\
 +&& \omega_0 (1-n_1)(n_{i+1}-n_i) + \omega_1 n_1 (n_{i+1}-n_i)\,]. 
 \end{eqnarray}

For the case $i = 1$, we need only condition on $n_1$ and $n_2$. We define the function
\[ g(n_1,n_2) = \frac{d}{dt} \prob{n_1(t)=1| n_1, n_{2}} = \frac{d}{dt} \expect{n_1(t)=1| n_1, n_{2}}.\]
As before, $g(0,n_2)$ gives the instantaneous rate of the jump $n_1 = 0 \to n_1=1$, while $-g(1,n_2)$ is the rate of the jump $n_1 = 1 \to n_1=0$. The values of $g(n_1,n_2)$ are enumerated in Table \ref{tab:enumeration} (right). It is evident that
\[  g(n_1,n_2) = \omega_\mathrm{a} c (1- n_1) - \omega_\mathrm{d} n_1 \nonumber + \omega_\mathrm{h} (n_2 - n_1) + \omega_0 (1-n_1)(n_2-n_1) \nonumber +\omega_1 n_1(n_2-n_1), \]
for all $(n_1, n_2) \in \{0,1\}^2$. Thus
\begin{eqnarray*} \frac{d}{dt} \expect{n_1} =&& \expect{\frac{d}{dt} \expect{n_1(t)=1| n_1, n_{2}}} = \expect{g(n_1, n_2)} \\
=&& \expect{\omega_\mathrm{a} c (1- n_1) - \omega_\mathrm{d} n_1 \nonumber + \omega_\mathrm{h} (n_2 - n_1) + \omega_0 (1-n_1)(n_2-n_1) \nonumber +\omega_1 n_1(n_2-n_1)}, 
\end{eqnarray*}
which is the expression given in equation 2 of the main text.

\section{Derivation and Solution of the Mean-Field Model}
For brevity, we return to the notation $\langle \,\cdot \, \rangle = \expect{\,\cdot \,}$ used in the main text. Starting with equation \ref{disc}, we apply a mean-field approximation $\langle n_1 n_i \rangle \approx \langle n_1 \rangle \langle n_i \rangle$, which neglects correlations between $n_1$ and the other $n_i$. This gives
\begin{eqnarray} \label{mfdisc}
  \frac{d \langle n_i \rangle}{dt} = &&\omega_\mathrm{a} c \left(1- \langle n_i \rangle\right) - \omega_\mathrm{d} \langle n_i \rangle + \omega_\mathrm{h} \left( \langle n_{i+1} \rangle - 2 \langle n_i \rangle + \langle n_{i-1} \rangle \right) \nonumber \\
  &&+ \left[\omega_0 \left(1-\langle n_1 \rangle\right) + \omega_1 \langle n_1 \rangle \right] \left(\langle n_{i+1} \rangle -\langle n_i \rangle\right).
\end{eqnarray}
Let $b$ be the width of the binding site on the polymer. Define $x = b(i-1)$, which is the distance from the shrinking tip along the length of the polymer. Also, define $u(x,t) = \langle n_{x/b+1} \rangle (t)$, which is proportional to the expected linear density of the protein at point $x$. Under these definitions, it follows from equation \ref{mfdisc} that
\begin{eqnarray*}
  \partial_t u(x,t) = &&\omega_\mathrm{a} c (1- u(x,t)) - \omega_\mathrm{d} u(x,t) + \omega_\mathrm{h} b^2 \left( \frac{u(x+b,t) - 2 u(x,t) + u(x-b,t)}{b^2} \right) \nonumber \\
  &&+ \left[\omega_0 b (1-u(0,t)) + \omega_1 b u(0,t)\right]\left(\frac{u(x+b,t)-u(x,t)}{b}\right). 
\end{eqnarray*} 
Making $x$ continuous and taking the limit $b \to 0$, we obtain the nonlinear advection-diffusion equation
\begin{equation}\label{dimensionfulpde}
\partial_t u =  D \partial_x^2 \, u + [v_0(1-u(0,t)) + v_1\,u(0,t)] \, \partial_x u + k_\mathrm{a} b\, c (1-u) - k_\mathrm{d} \,u,\end{equation}
where we have identified the following macroscopic, experimentally measurable parameters: $D$, the one-dimensional diffusion coefficient of the protein; $v_0$, the natural shrinkage velocity of the polymer in absence of the protein; $v_1$, the shrinkage velocity of the polymer when it is saturated with the protein; $k_\mathrm{a}$, the protein attachment rate per unit concentration and per unit length of polymer; and $k_\mathrm{d}$, the detachment rate of the protein. The relationship between these macroscopic parameters and the microscopic parameters is given in Table \ref{tab:params}. 

\begin{table}[b]
\renewcommand{\arraystretch}{1.6}
    \centering

    \begin{tabular}{@{\quad}c@{\quad}c@{\quad}c@{\quad}c@{\quad}}
\hline \hline
Process    &  Microscopic Parameter  &  Macroscopic Parameter &  Dimensionless Parameter \\ \hline
Diffusion  & $\omega_\mathrm{h}$              & $D = b^2 \omega_\mathrm{h}$     & $\Dbar = \omega_\mathrm{h}/\omega_0$ \\ \hline
Attachment & \begin{tabular}[c]{@{}c@{}}$\omega_\mathrm{a}$\\ $c$\end{tabular} & \begin{tabular}[c]{@{}c@{}}$k_\mathrm{a} = \omega_\mathrm{a} /b$\\ $c$\end{tabular} & \multirow{2}{*}{\begin{tabular}[c]{@{}c@{}}$\cbar = \left(\omega_\mathrm{a}/\omega_\mathrm{d} \right) c$\\ $\kbar = \omega_\mathrm{d}/\omega_0$\end{tabular}} \\ \cline{1-3}
Detachment & $\omega_\mathrm{d}$              & $k_\mathrm{d} = \omega_\mathrm{d}$ & \\ \hline
Shrinkage  & \begin{tabular}[c]{@{}c@{}}$\omega_0$\\ $\omega_1$\end{tabular} & \begin{tabular}[c]{@{}c@{}}$v_0$\\ $v_1$\end{tabular} & $\vbar = \omega_1/\omega_0$ \\\hline \hline
\end{tabular}

    \caption{Relationship between the microscopic, macroscopic, and dimensionless parameters.}
    \label{tab:params}
\end{table}
It is convenient to work with the following dimensionless version of equation \ref{dimensionfulpde},
\begin{equation} 
    \partial_{\tbar} \rho = \Dbar \, \partial_{\xbar}^2 \rho + [1-(1-\vbar) \rho(0,\tbar)]\, \partial_{\xbar} \rho + \kbar \cbar (1 - \rho) - \kbar \rho,  \label{pde}
\end{equation}
where $\xbar = x/b$, $\tbar = t\omega_0$, and $\rho(\xbar,\tbar) = u(b\, \xbar, \tbar/\omega_0)$. The dimensionless parameters $\Dbar$, $\kbar$, $\cbar$ and $\vbar$ are defined in Table \ref{tab:params}. As discussed in the main text, the boundary conditions are
\begin{subequations}
\begin{align}
  &\lim_{\xbar \to \infty} \rho(\xbar, \tbar) = \rhoeq = \frac{\cbar}{1+\cbar}, \\
  &-\Dbar\, \partial_{\xbar}\rho(0,\tbar) - [1-(1-\vbar) \rho(0,\tbar)]\, \rho(0, \tbar) =  \kbar \cbar (1 - \rho(0,\tbar)) - (\kbar + \vbar) \rho(0,\tbar).
  \end{align}
\end{subequations}

With these boundary conditions, the stationary solution to equation \ref{pde} has the form
\begin{equation*} \label{statSol}
    \rho(\xbar) = (\rho(0)-\rhoeq) e^{-\lambda \xbar} + \rhoeq,
\end{equation*}
where
\[\lambda = \frac{1-(1-\vbar)\rho (0) + \sqrt{4 \Dbar \kbar(1+\cbar)+\left[1-(1-\vbar)\rho (0)\right]^2}}{2 \Dbar},\]
and $\rho(0)$ is the solution to
\begin{equation} \label{poly} A_0 + A_1 y + A_2 y^2 + A_3 y^3 = 0, \qquad y \in [0,1],\end{equation}
where
\begin{eqnarray*}
    A_0 &&= \cbar^2 \kbar (1 - \Dbar + \kbar + \cbar \kbar), \\
    A_1 &&=  \cbar (1-\vbar-2 \kbar (-\Dbar+\kbar+\vbar)-\cbar \kbar (1-2 \Dbar+2 (2+\cbar) \kbar+\vbar)),\\
    A_2 &&=  -\kbar-\Dbar \kbar+\cbar^3 \kbar^2+2 \cbar (-1+\vbar)-\vbar+(\kbar+\vbar)^2+\cbar^2 \kbar (-1-\Dbar+3 \kbar+2 \vbar)+\cbar \kbar (-2-2 \Dbar+3 \kbar+4 \vbar),\\
    A_3 &&= (1 - \vbar) (\cbar + \kbar + \cbar (2 + \cbar) \kbar + \vbar)
\end{eqnarray*}
Numeric inspection suggests that for any positive $\Dbar, \kbar, \cbar$ and $\vbar$, the polynomial in equation \ref{poly} has exactly one root in the interval $[0,1]$, and so $\rho(0)$ is uniquely determined. This root can be expressed analytically using the cubic formula, but the resulting expression is lengthy and opaque. The solution simplifies considerably in certain limiting cases. If $\kbar \to 0$ (equivalently, $\omega_\mathrm{d} \gg \omega_0$), we have
\[\rho(0) = \frac{\cbar}{\cbar+\vbar}, \qquad \qquad \lambda = \frac{(\cbar+1) \vbar}{(\cbar+\vbar) \Dbar}. \]
If $\Dbar \to 0$ (or $\omega_\mathrm{h} \ll \omega_0$), a discontinuity appears at the tip as $\lambda$ diverges. The density for all $\xbar>0$ is $\rhoeq$, and at $\xbar=0$ it is
\begin{equation*} 
\rho(0) = \frac{\cbar (\kbar \cbar+\kbar+1)}{\kbar (\cbar+1)^2+\cbar+\vbar} 
\end{equation*}

\section{Simulation and numerical methods}
\subsection*{Stochastic lattice-gas simulation}
Simulations were carried out on a finite lattice. Because the lattice shrinks, it is necessary for the initial length $L_0$ to be large enough that the protein dynamics stationarize before finite size effects become significant. The simulation method is similar to Gillespie's algorithm for Markov jump processes and has the following steps
\begin{enumerate}
    \item Initialize a boolean vector $\mathbf{m} \in \{0,1\}^{L_0}$ drawn from the initial distribution $\prob{n_i=1} = \rhoeq$. The integer $L$ marks the position of the tip and decreases over time as the lattice shrinks (we are no longer working in the frame of the tip). Initialize $L=L_0$ and $t=0$. At any point in time, the state of the system is determined by the first $L$ entries of $\mathbf{m}$, which represents the portion of the polymer that has yet to shrink away at that time.
    \item Count the number of possible hops $N_h$, attachments $N_a$, and detachments $N_d$ within the first $L$ entries of $\mathbf{m}$.
    \item The instantaneous transition rate to the next state is $\omega = \omega_\mathrm{h} N_h + \omega_\mathrm{a} c N_a + \omega_\mathrm{d} N_d + \omega_0 (1-m_L) + \omega_1 m_L $. Draw an $\mathrm{Exp}(\omega)$ random variable, and add it to $t$.
    \item Choose whether the next transition is a hop, an attachment, a detachment, or a lattice shrinkage event. The probabilities of these choices are $\omega_\mathrm{h} N_h /w$ for a hop, $\omega_\mathrm{a} c N_a /\omega$ for an attachment, $\omega_\mathrm{d} N_d/\omega$ for a detachment, and $(\omega_0 (1-m_L) + \omega_1 m_L)/\omega$ for a shrinkage event.
    \item If the next transition is a hop, then there are $N_h$ possibilities for the next state. Pick one of these possibilities with uniform probability, and update $\mathbf{m}$ accordingly. If the next transition is an attachment, fill one of the $N_a$ vacant sites, and if it is a detachment, set one of the $N_d$ occupied sites to zero. If it is a lattice shrinkage event, update $L$ to $L-1$.
    \item Return to step 2.
\end{enumerate}
This algorithm was implemented in the Julia programming language \cite{Julia-2017}. The source code is available upon request.

\begin{figure}[t]
    \centering
    \includegraphics[scale=0.97]{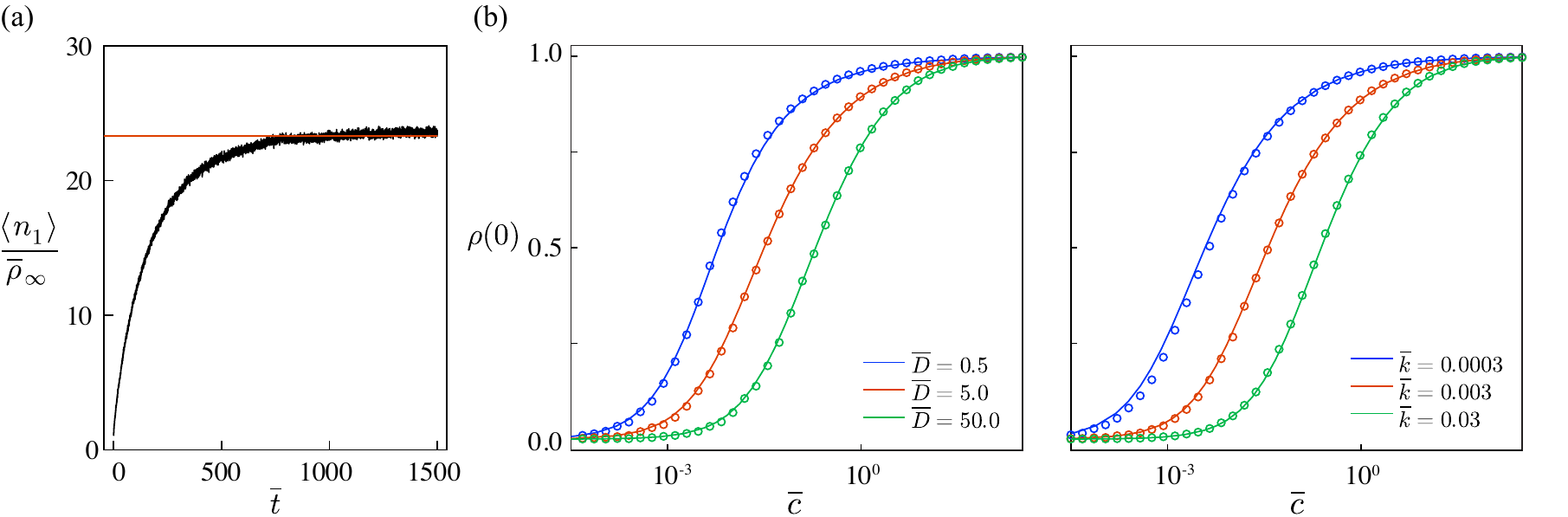}
    \caption{Agreement between the Lattice-Gas and Mean-Field Models in the steady-state. (a) Monte-Carlo simulations of the lattice-gas model (black curve) predict that the fold-enrichment at the tip $\langle n_1 \rangle/\rhoeq$ increases over time up to a steady-state value. Orange line is the steady-state predicted by the mean field model. (b) Steady-state tip occupancy $\rho(0) = \langle n_1 \rangle$ as a function of protein concentration $\cbar$ at different values of $\Dbar$ (left) and $\kbar$ (right). Circles are Monte-Carlo estimates from the lattice-gas model. Solid curves are solutions of the mean-field model.}
    \label{fig:stochfig}
\end{figure}

Monte Carlo estimates of the mean lattice occupancies were calculated from an ensemble of simulations. As seen in Figure 2 of the main text, the occupancy profile approaches a steady-state that is well predicted by the mean-field model. This is further detailed in Figure \ref{fig:stochfig}a, which shows that the mean tip occupancy $\langle n_1 \rangle$ gradually approaches the stationary value predicted by the mean-field model. The steady-state tip occupancies predicted by the lattice-gas and mean-field models are in excellent agreement over the full range of concentrations $\cbar$ and for several values of $\kbar$ and $\Dbar$ (Figure \ref{fig:stochfig}b). 

\subsection*{Numeric integration of mean-field model}
A nonlocal finite-difference scheme was implemented for the numeric integration of equation 6 of the main text. Space is partitioned into a mesh $i=1, 2, \dots, N$. Imposing the boundary conditions in the main text, the discretized equation 6 takes the form
\begin{align*}
\dot{\rho_1} &= D \frac{(\rho_2-\rho_1)}{\Delta x^2} + (1-\rho_1)\frac{\rho_2}{\Delta x} \\
\dot{\rho_i} &= D \frac{(\rho_{i+1}-2\rho_i+\rho_{i-1})}{\Delta x^2} + (1-\rho_1) \frac{(\rho_{i+1}-\rho_i)}{\Delta x}, \qquad  (2\leq i \leq N-1) \\
\rho_N &= D \frac{(-2\rho_N+\rho_{N-1})}{\Delta x^2} + (1-\rho_1)\frac{-\rho_N}{\Delta x}
\end{align*}
These ODEs were solved via the TR-BDF2 algorithm in the Julia programming language using the DifferentialEquations.jl package.
\section{Detailed Experimental Methods}
\subsection*{Spastin Preparation}
The GFP-spastin construct used in this study contains a StrepII-tag (WSHPQFEK) at its N-terminus, followed by a superfolder Green Fluorescencent Protein
(sfGFP) fluorophore, and then the full-length short isoform of \textit{Drosophila melanogaster} spastin (208 aa-end). A flexible linker (GGSGGGSGGGS) was used to connect spastin and the sfGFP fluorophore. A PreScission protease cleavage site (LEVLFQGP) was introduced between the StrepII-tag and sfGFP during cloning. Our StrepII-sfGFP-spastin construct was prepared from the previously described pET-His6-MBP-spastin construct \cite{Kuo2019a} through two cloning steps. First, the MBP tag was replaced by a PCR fragment containing sfGFP and the N-terminal PreScission protease cleavage site using HiFi assembly (NEB). The His6-tag was then replaced by a StrepII-tag using Q5 site-directed mutagenesis (NEB). All constructs were validated by DNA sequencing.

The StrepII-sfGFP-spastin construct was expressed in \textit{Escherichia coli} (Rosetta(DE3) competent cells; Novagen) overnight at 16 °C. Cell pellets were resuspended in cold lysis buffer (30 mM Hepes, pH 7.4, 0.3 M NaCl, 5\% glycerol, 1 mM DTT, 10 \textmu M ATP) supplemented with 0.1 mg/mL lysozyme, 0.3 U/\textmu L bezonase, and protease inhibitors (0.2 mM pefabloc, 5 \textmu g/mL leupeptin). The cells were lysed by sonicating the suspension on ice. The soluble portion of the lysate was loaded onto an affinity column (Strep-Tactin Superflow Plus; QIAGEN), washed with lysis buffer, and then eluted with  2.5 mM D-desthiobiotin in lysis buffer. Collected fractions were pooled and concentrated using a centrifugal filter (Amicon Ultra 50 KDa; Millipore). The solution was then loaded onto a size-exclusion column (Superdex 200; GE-Healthcare). Collected fractions were pooled, concentrated, flash-frozen, and stored at $-80$\textdegree C.

For the shrinkage rate measurements (Figure 4, main text), a non-fluorescent His\textsubscript{6}MBP-tagged spastin construct was used. This construct has a lower tendency to aggregate, allowing the protein concentration to be tuned more reliably. The cloning and purification procedures are detailed in \cite{Kuo2021}.

\subsection*{Microtubule Dynamic Assay}
The microtubule dynamics assay used here is a modification of the protocol detailed in \cite{krismethods}. Tubulin was purified from bovine brains as described in \cite{Castoldi2003}.  A microfluidic reaction chamber was prepared consisting of 5-Polydimethylsiloxane (PDMS) microchannels fixed on a cover glass. These reaction chambers allow reagents to be rapidly exchanged during imaging. The microchannels were prepared using a silicon master mold that produces channel dimensions of 500 \textmu m$\times$1 cm$\times$100 \textmu m.  The cover glasses used to prepare the chambers were cleaned and silanized as described in \cite{Gell2010}. 

GMPCPP-stabilized microtubule seeds were grown as described in \cite{Gell2010} using biotinylated tubulin (Cytoskeleton Inc.) diluted with unlabelled tubulin to a 5\% labelling stoichiometry. The seeds were immobilized onto the silanized surface within the flow channels using anti-biotin antibody (Sigma-Aldrich B3640). Following this, dynamic GTP/GDP tubulin ``extensions" were grown by incubating with 12 \textmu M unlabeled tubulin, 1 mM GTP, and 5 mM DTT in BRB80 buffer. All reactions were carried out at 28\textdegree C.

\subsection*{Imaging spastin on microtubules}
Purified StrepII-sfGFP-DmSpastin was visualized on microtubules via Total Internal Reflection Fluorescence (TIRF) microscopy as described in \cite{Gell2010}. A 488nm excitation laser was used. The microscope exposure time was set to 100 ms. The frame rate was 9.8 Hz and the laser intensity was about 0.05 kW cm$^{-2}$. All image analysis was performed using Fiji \cite{Schindelin2012} and JuliaImages \cite{Julia-2017}.

The imaging buffer consisted of BRB80 supplemented with 50 mM KCl, 1 mM MgCl2, and 10 mM DTT, as well as 1 mM AMP-PNP, which is a nonhydrolyzable ATP-analogue that binds spastin but does not allow it to sever microtubules. To slow photobleaching, we also included an oxygen-scavenging reagent mix in the imaging buffer (40 mM glucose, 40 mg/mL glucose oxidase, 16 mg/mL catalase, and 0.1 mg/mL casein). To visualize spastin's tip enrichment, we grow dynamic microtubules from seeds as described above. We introduce a solution containing 50 nM GFP-Spastin, 8 \textmu M unlabeled tubulin, and 1 mM GTP in imaging buffer. We then begin imaging and immediately replace the solution with one containing just 50 nM GFP-Spastin in imaging buffer. The removal of tubulin causes microtubules to begin shrinking. The washout assay is carried out similarly, but without including spastin in the second solution. 

Microtubules were visualized label-free using Interference-Reflection Microscopy (IRM) as described in \cite{Mahamdeh2018}. For simultaneous visualization of spastin via TIRF and microtubules via IRM (e.g. Figure \ref{fig:accum}), we used an image splitter (Teledyne-Photometrics) to spectrally and spatially separate TIRF and IRM images on two halves of the same camera sensor (described in \cite{Tuna2022}). This allows simultaneous high-speed imaging of spastin and shrinking microtubules. 
\subsection*{Measurement of microtubule shrinkage velocity}
For measurements of the microtubule shrinkage velocity (Figure 4, main text), non-fluorescent His\textsubscript{6}MBP-tagged spastin was used. The imaging buffer was BRB80 supplemented with 1 mM AMP-PNP. Dynamic microtubules were grown as described above, with one modification: in the ``extension stage", the unlabelled spastin was included in the solution at the appropriate measurement concentration. Then, the solution was replaced with one containing only spastin and no tubulin. The shrinkage of the microtubules was visualized via IRM. The steady-state shrinkage velocity was determined from kymographs of the shrinkage process.
\subsection*{Estimation of landing rate, dwell time, and diffusion coefficient}
To estimate the on-rate, we observed via TIRF the landing of single fluorescent spastin molecules on GMPCPP seeds. The number of landing events during a fixed period was counted to estimate the landing rate. To estimate dwell times and diffusion coefficients, a washout assay was performed with an initial GFP-spastin concentration of 100 nM. This allowed the visualization and tracking of single molecules. Molecules were tracked using the TrackMate plugin in Fiji \cite{Tinevez2017}. The diffusion coefficient was extracted from the variance of the step size distribution after subtracting the variance due to localization error.

\section{Supplementary Experimental Results}
\begin{figure}
\begin{minipage}{0.48\linewidth}
    \includegraphics[page=1, scale=1]{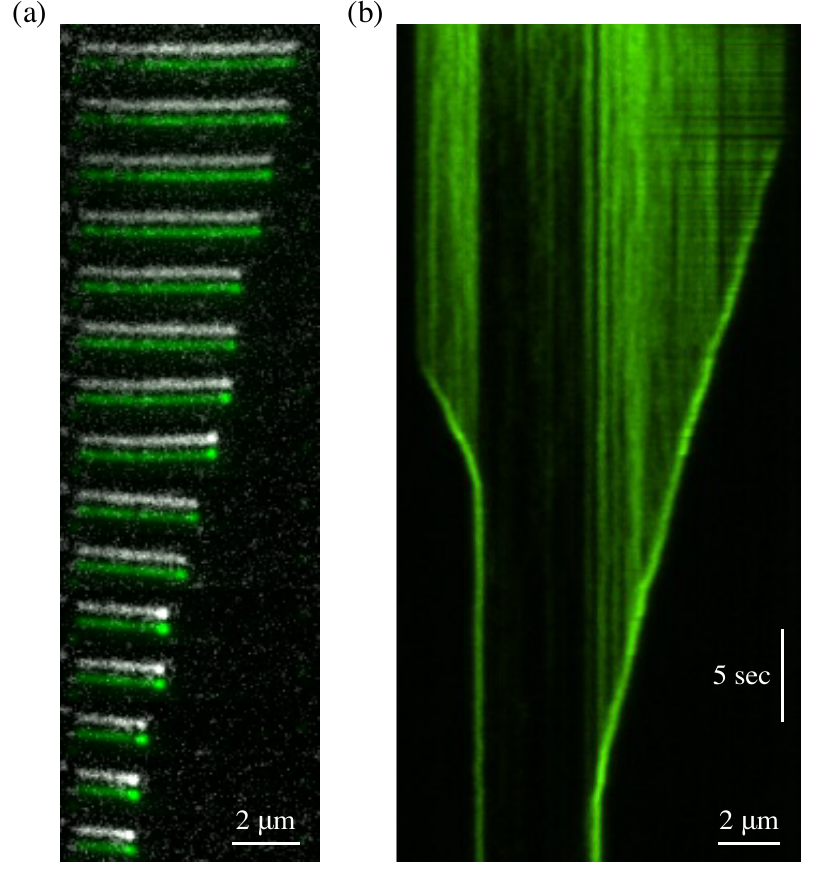}
    \caption{Accumulation of spastin on shrinking microtubule ends. (a) Shrinking microtubules (white) and GFP-spastin (green) were simultaneously visualized via IRM and TIRF, respectively. A representative time series is shown. The TIRF and IRM channels are shifted apart vertically by 8 pixels to facilitate comparison. Consecutive snapshots are separated by 10.2 sec. (b) Kymograph from a washout assay showing that spastin's shrinking end enrichment occurs on both plus and minus ends.}
    \label{fig:accum}
\end{minipage} \hfill
\begin{minipage}{0.48\linewidth}
    \includegraphics[page=2,scale=1]{Figure_S2_and_S3-SuppExperiments.pdf}
    \caption{Diffusion of GFP-spastin on GMPCPP-stabilized microtubules. 100 nM spastin was added to immobilized GMPCPP-stabilized microtubules in imaging buffer (containing 1 nM AMP-PNP). The spastin was washed out with imaging buffer. The particles that dwelled on the microtubules after washout were visualized via TIRF, and many of these were diffusive. Two sample trajectories are shown.}
    \label{fig:diffu}
\end{minipage}
\end{figure}

To verify that the observed region of spastin enrichment coincides with the position of the shrinking microtubule end, we simultaneously visualized the microtubule (via IRM) and the GFP-spastin (via TIRF). A sample time series is shown in Figure \ref{fig:accum}a. It is evident that, up to the resolution limits of the microscope, the enriched region of spastin coincides with the shrinking end. End enrichment of spastin is also visible in the IRM channel of Figure \ref{fig:accum}a.

In Figure \ref{fig:accum}b, a sample kymograph is provided showing that spastin is enriched both on shrinking plus ends and shrinking minus ends. This behavior was evident in all microtubules observed ($n>100$).

\begin{figure}
    \centering
    \includegraphics[scale=1]{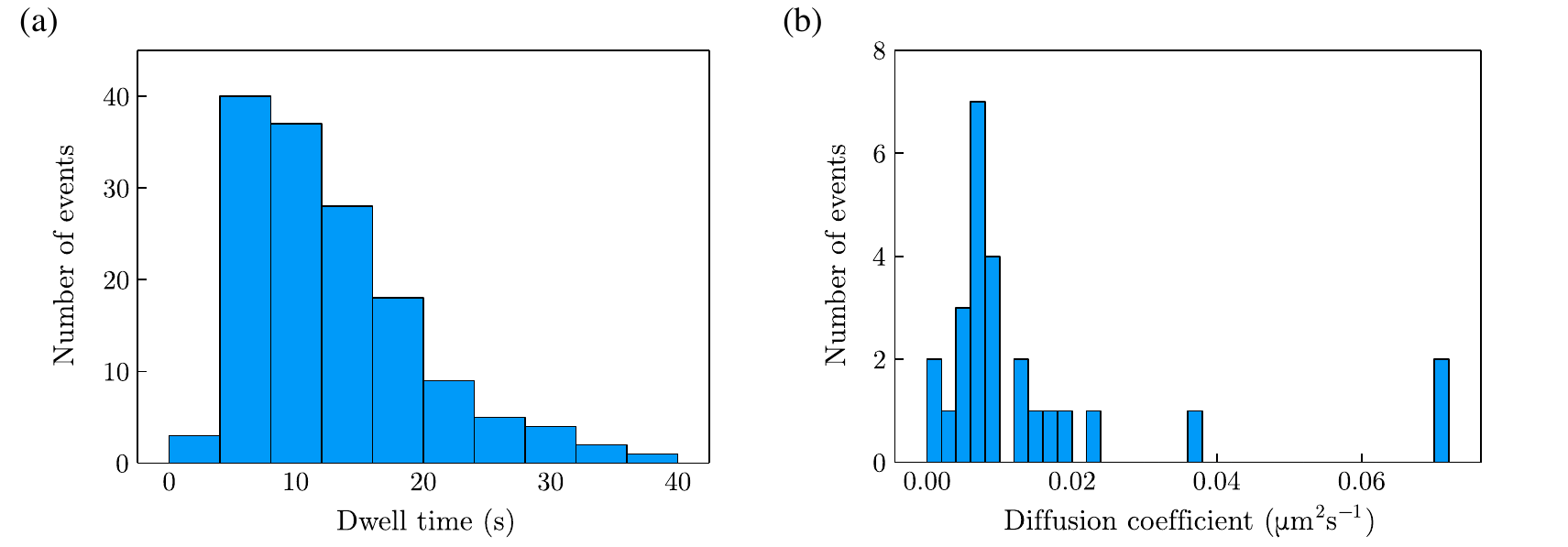}
    \caption{Trajectory statistics of single GFP-spastin particles on microtubules. (a) Distribution of particle dwell times, with an estimated mean of $(17.7 \pm 1.4) \mathrm{s}$. This histogram is not corrected for photobleaching; the bleaching time was estimated from background puncta at $\approx \!\!130$s, so the dwell time of spastin is likely underestimated by 10-20\%. (b) Distribution of diffusion coefficients estimated from the step size distributions. Only estimates with a relative error below 50\% are included.}
    \label{fig:dwellAndDiff}
\end{figure}

Example single-particle trajectories are shown in Figure \ref{fig:diffu}. The dwell time distribution of the trajectories is shown in Figure \ref{fig:dwellAndDiff}a. The mean dwell time was estimated using the estimator developed in \cite{Deemer1955} in order to compensate for the undersampling of very short and very long trajectories. The estimated mean is $(17.7 \pm 1.4) \mathrm{s}$. To ensure that the measured lifetimes represent detachment of spastin from the microtubule and not photobleaching of the fluorophore, we estimated the bleaching rate by measuring the intensity of surface-bound GFP-spastin aggregates in the same experiment. The bleaching time $\tau_\mathrm{B}$ is estimated by fitting the intensity time series to a function of the form $A \exp(-t/\tau_\mathrm{B})+C$. This yielded an estimate of $\tau_\mathrm{B} \approx 130 \mathrm{s}$, which is significantly longer than the observed dwell times. The true spastin dwell time is therefore likely to be only 10-20\% greater than our estimated mean dwell time.

\clearpage
\bibliography{bibliography}